\centerline{\bf On Quantum Contributions to Black Hole Growth}
\medskip
\centerline{M.\ Spaans}
\smallskip
\centerline{Kapteyn Astronomical Institute, University of Groningen, The Netherlands; spaans@astro.rug.nl}

\medskip\medskip
\noindent {\bf Abstract}
\smallskip
The effects of Wheeler's quantum foam on black hole growth are
explored from an astrophysical perspective.
Quantum fluctuations in the form of mini ($10^{-5}$ g) black holes can couple
to macroscopic black holes and allow the latter to grow exponentially in mass
on a time scale of $\sim 10^9$ years.
Consequently, supermassive black holes can acquire a lot of their mass
through these quantum contributions over the life time of the universe.
This alleviates the need for very efficient forms of baryonic matter
accretion more recent than a redshift $z\sim 6$.
Sgr A$^*$ in the Milky Way center is a candidate to verify this quantum
space-time effect, with a predicted mass growth rate of $4\times 10^{-3}$
M$_\odot$ yr$^{-1}$.
A few comments on the possibility and consequences of dark matter as quantum
grown black holes are made, with a big crunch fate of the universe.

\medskip\medskip
\noindent {\bf Introduction}
\smallskip
Almost every galaxy is known to harbor a supermassive black hole (BH) in its
nucleus, with masses ranging from $10^5$ to more than $10^{10}$
M$_\odot$. Best known is the Magorrian relationship (Magorrian et al.\ 1998),
which states that the mass of the central BH is
$\sim 10^{-3}$ of the stellar bulge mass, albeit with a large scatter.
Furthermore, BHs with masses as large as $10^9$ M$_\odot$ have been detected
around a redshift of $z\sim 6$ (Fan et al.\ 2001). This leaves about a
billion years to grow such monsters. In any case, the growth of BHs (be they
large or small) is of fundamental astrophysical importance.

In recent years, many efforts have been devoted to understand the growth of
BHs, both today and in the early universe (e.g., Aykutalp et al.\ 2013; Park \&
Ricotti 2012; Kim et al.\ 2011; Di Matteo et al.\ 2005; Wada \& Habe 1995,
1992; and references therein).
It has become clear from these studies that feedback effects may render it
difficult to achieve an appreciable duty cycle for gas accretion. Nevertheless,
building up a massive BH at a fraction of Eddington seems possible.

That being said, any additional growth would certainly aid the occurrence of
present-day supermassive BHs.
This work explores the possibility that space-time fluctuations, as an
expression of quantum gravity (Wheeler 1957), allow the mass growth of BHs on
cosmologically interesting time scales (Spaans 1997, 2013a).
Such an unconventional growth mode operates in the absence of any matter and
therefore does not suffer from baryonic feedback effects. Some of its features
are quantified and discussed in this paper.

\medskip\medskip
\noindent {\bf Properties of Quantum Space-Time}
\smallskip
As one approaches the Planck scale of $l_P\sim 10^{-33}$ cm, space-time is
expected to deviate from a smooth structure in various manners.
Specifically, Wheeler (1957) has argued that space-time on the Planck scale
hosts a multitude of Planck mass ($m_P=10^{-5}$ g) mini BHs that pop out off
and back into the vacuum every Planck time $t_P=l_P/c$ (for the speed of
light $c$).
This because Einstein gravity is not invariant under conformal, so local scale
changing, transformations.
Such a form of microscopic space-time is called quantum foam and is known
to be quite unstable (1987).
In Spaans (1997) it was found that a stable quantum foam results globally
because only macroscopic (so long-lived) BHs induce pairs of mini BHs each
Planck time.
The fluctuations that Wheeler's quantum foam embody are thus a miniature of
the global distribution of BHs (Spaans 2013b).
Wheeler's quantum foam then also couples to macroscopic BHs and this allows
them to gain mass if specific early universe conditions pertain. The reader
is referred to Spaans (1997, 2013a) for a detailed description of quantum
space-time, but some of the arguments are described below.

The main idea is that many distinct paths weave the fabric of space-time and
allow the implementation of the Feynman path integral.
However, the Planck scale is characterized by large fluctuations in shape
that render the identification of such paths ill defined. Topology
(connectivity) is therefore essential because it allows one to implement
paths as an expression of a multiply connected space-time, whatever its
geometry. That is, two paths are distinct when they cannot be obtained from
each other by a continuous deformation of space-time.
Furthermore, to identify a path one needs a proper example to compare to in
the first place. Consequently, no topologically distinct path can exist
individually under the quantum act of observation.
I.e., topological fluctuations on $l_P$ take the form of multiple copies of
any path through quantum space-time.

This implies that a long-lived ($>>t_P$) BH induces pairs of mini BHs.
The presence of a BH horizon in four-space should be identifiable at times
separated by about $\delta t=t_P$. This because every BH has a thermodynamic
temperature and can re-radiate absorbed mass-energy. The latter constitutes
a path that enters and exists the horizon at separate times, equivalent to a
wormhole topology. Also, no local or global continuous transformation of
space-time can deform this path into one that does not pass through the
horizon, because that would remove it and change space-time connectivity.
This renders BH effects distinguishable from other space-time regions in a
topological sense.

To establish this path for pairs of moments spanning
$\delta t$, it is necessary to zoom in on the BH horizon to very small scales.
In fact, one must scrutinize the horizon with about a Planck mass worth of
energy ($h/t_P\sim m_Pc^2$ for Planck's constant $h$) within a region of size
$l_P=ct_P$. This amounts to the creation of mini BHs (Wheeler 1957, 1987) that
are topologically identical to the long-lived BH.
Hence, quantum uncertainty in the Planck-scale structure of a BH horizon
induces space-time fluctuations in the form of mini BH pairs, every Planck
time. These mini BHs acquire their mass from the huge Planckian vacuum
energy and not the BH itself. In an expanding universe energy conservation
does not hold globally because there is no time translation invariance.

In order to quantify the possible macroscopic consequences of this quantum
gravity effect, Spaans (2013a) explored the subtleties of embedding BHs, and
thus their induced mini BHs, into space-time.
The four-dimensional nature of BH horizons requires a global embedding of
the universe, so one that connects time intervals $>>t_P$ and spatial scales
$>>l_P$ through four-space, if BHs have longevity under Hawking (1974)
evaporation.
It turns out that four-space is partitioned into time slices with a width of
$t_X\approx 4\times 10^{12}t_P$ and volumes $L_f^3$ with
$L_f\approx 2\times 10^{14}$ cm.

The first BH that has a life time under Hawking evaporation longer than
the contemporary age of the universe sets $t_X$, which scales as BH mass cubed.
This BH $X$ has a mass $M_X\sim 0.3$ g at time $t_I\sim 10^9t_p$, when BHs
are on average $m_X < 0.03$ g with life times $< t_I$. $M_X$ is
an excursion from that mean (see the Discussion). $m_X$ is $< 10^3m_P$
on thermodynamic grounds because the Planck energy $10^{19}$ GeV exceeds the
grand unified theory (GUT) energy $10^{16}$ GeV by a factor $\sim 10^3$.
Before $t_I$ the universe goes through inflation and a GUT phase transition.

The first occurrence of a local time scale $t_X>t_I$ necessitates a global
four-space embedding to accommodate this BH as a proper quantum history
across a time slice of width $t_X>t_I$.
When BH $X$ causes the embedding that allows its evolution as a part of the
wave function of the universe, it also imposes a temporal spread $t_X>>t_P$ on
the mini BH fluctuations that identify its horizon. Afterall, any object with
a longevity $t_X$ has a correspondingly diffuse temporal presence in
four-space.
The time scale $t_X$ holds for all later time BHs because the creation of BH
$X$ is a permanent part of the universe's past.
I.e., BH X allows a 4-space topology in the sense of Mach. So one that has
longevity in a global sense and also pertains to mass in the form of BHs:
global topology and geometry determine the changes in Planck scale topology
and local motions of matter. The time scale on which this occurs must be
tied to the first occurrence of a BH that is global in a temporal sense, so
that can live longer than the contemporary age of the universe at its
inception.

Through such an initial cause driven embedding a specific volume $L_f^3$
is globally frozen in as well.
Obviously, the spatial scale $L_f$ is the size of the entire universe at $t_I$.
It can be computed from today's number of macroscopic BHs,
$N_{\rm BH}(0)\approx 10^{19}$. This is the total number of
stellar BHs. It assumes a Salpeter initial mass function with 1\% of all stars
more massive than 8 M$_\odot$ and 10\% of all supernovae creating a BH.
I.e., 1 in $10^3$ stars yields a BH and there are $\sim 10^{11}$ galaxies,
each containing $\sim 10^{11}$ stars. Primordial BHs are neglected
and the observable universe is a proxy.
This value links $L_f$ to the current vacuum (dark) energy density
$\Lambda (0)\approx 10^{-29}$ g cm$^{-3}$, carried by Wheeler's
quantum foam, through $\Lambda (0)=2m_PN_{\rm BH}(0)/L_f^3$.
So $L_f$ is the spatial scale to which induced mini BHs couple.
In all, the formation of BH $X$ forever endows four-space with a global
(Machian) notion of quantum uncertainty for the temporal
($t_X\approx 4\times 10^{12}t_P$) and spatial ($L_f\approx 2\times 10^{14}$ cm)
embedding of mini BHs.

\eject
\noindent {\bf Results for Macroscopic BHs}
\smallskip
Given that $t_X>>t_P$ and $L_f>>l_P$ together constitute a local perturbative
limit, it is possible to compute how strongly mini BHs and macroscopic BHs
interact.
$L_f$ represents how any time slice through four-space of thickness $t_X$ is
sub-partitioned, and $L_f$ amounts to $A=L_f/l_P\approx 2\times 10^{47}$ Planck
lengths.
This implies a dimensionless coupling of $A^{-1}\approx 5\times 10^{-48}$
between mini BHs and a local macroscopic BH horizon generating them. The
crucial point here is that our universe has expanded since $t_I$ to a size
many times larger than the Machian scale $L_f$. So that the latter is now a
local volume that quantum fluctuations couple to.
That is, quantum partitioning lowers the probability to find these mini BHs
within $l_P$ of the horizon to $(4\pi R_s^2l_P)/(4\pi R_s^2L_f)=A^{-1}$,
independent of the Schwarzschild radius $R_s$ and thus the BH mass.
Furthermore, all induced mini BHs are spread across a time slice $t_X>>t_P$.
Consequently, the time scale $t_Q=At_X\approx 10^9$ years quantifies how long
it takes a macroscopic BH to build up an order unity interaction with its
reservoir of mini BHs. This as the BH evolves over time intervals
$\Delta t>>t_X$.

The quantum foam expresses space-time fluctuations in the form of mini BHs due
to the lack of conformal invariance in general relativity, specifically for BH
horizons in this work. So $t_Q$ determines how rapidly perturbations in the
Schwarzschild radii $R_s$ of macroscopic BHs, as a result of mergers with mini
BHs, grow conformally. I.e., one has a change in horizon scale
$$\Delta R_s=\Delta t R_s/t_Q.\eqno(1)$$
Because $R_s\propto M_{\rm BH}$, for BH mass $M_{\rm BH}$, one finds the
infinitesimal mass evolution
$$\Delta M_{\rm BH}/\Delta t=M_{\rm BH}/t_Q.\eqno(2)$$
This yields a quantum accretion rate $R_Q$ of
$$R_Q(t)=M_{\rm BH}(t)/t_Q\ \ \ {\rm M}_\odot \ {\rm yr}^{-1}.\eqno(3)$$
For times longer than $t_Q$, BHs significantly feed off their quantum foam
and gain mass. Thus, BHs spontaneously lower their thermodynamic temperature
$T_{\rm BH}\propto 1/M_{\rm BH}$ and become colder.

The fact that the interaction strength $t_p/t_Q\sim 10^{-60}$ is tiny,
highlights the intrinsic weakness of this autonomous growth (AG).
So when viewed as an instability, it is of very modest importance.
However, our universe has survived for so long that Wheeler's induced quantum
foam is pertinent in a cumulative sense.
AG does not require any ad hoc external field(s) from which BHs accrete.
Interestingly, $t_Q\approx 10^9$ years is close to the Eddington time of 0.45
billion years. This renders AG a cosmologically pertinent mechanism.

AG alone, so without mass accretion, follows
$$M_{\rm BH}(t\ge t_1)=M_1 e^{(t-t_1)/t_Q},\eqno(4)$$
with $t_1$ the time at which a seed BH is formed and $M_1$
its mass. The current age of the universe is about 14 billion years
($\sim 10^{61}t_P$). Hence, the
boost factor $B\equiv e^{t/t_Q}$ can be as large as $\sim 10^6$, thus allowing
for very significant AG. A fiducial time scale of 10 billion years, so
$z\sim 1.5$, still yields $B\approx 2\times 10^4$.

One expects stellar seed BHs of $1-30$ M$_\odot$ for
a zero or very low metallicity chemistry and direct collapse seeds of
$10^4-10^6$ M$_\odot$, both after a few hundred million years (e.g., Wise et
al.\ 2012; Klessen et al.\ 2012; Aykutalp \& Spaans 2011; Yoshida et al.\ 2008;
Spaans \& Silk 2006; Haiman 2006; Bromm \& Loeb 2003; Abel et al.\ 2002).
Therefore, a direct collapse seed of $10^5$ M$_\odot$ at redshift $z=20$
(140 million years after the big bang) quantum grows by a factor of
$\approx 8\times 10^5$ during the age of the universe and reaches the scale
of the biggest supermassive BHs known today ($>10^{10}$ M$_\odot$).
A Pop III seed BH of $3$ M$_\odot$ can grow from $z=10$ (400 million
years after the big bang) to a current mass of $2\times 10^6$ M$_\odot$.

The mass of our own Milky Way BH is about $4\times 10^6$ M$_\odot$
(Ghez et al.\ 2008, Gillessen et al.\ 2009). Hence, half of that mass can be
acquired through AG of a Pop III seed,
without the need for a massive direct collapse seed.
In fact, Sgr A$^*$ should grow by $\sim 4\times 10^{-2}$ M$_\odot$ in 10
years due to mini BHs.
A pulsar near the Milky Way central BH would be ideally suited to verify this,
by using it as a tracer of the time dependent BH gravitational potential.
One such object has been detected already (Kennea et al.\ 2013), and a
long-time monitoring program of its behavior would be useful.

\medskip\medskip
\noindent {\bf Discussion}
\smallskip
Supermassive BHs at high redshift benefit least from AG. This
because the characteristic time scale $t_Q$ allows for modest growth
(factor 2-3) upto $z\sim 6$.
This strengthens the need for massive seed BHs of $10^{4-6}$ M$_\odot$ at
redshifts beyond 10 and/or very efficient accretion onto Pop III seeds,
aided by (proto-)galaxy mergers.

On the other hand, the present-day Magorrian relationship is helped along.
AG attaches a minimal gain to the initial mass function of seed BHs,
irrespective of negative local feedback effects associated with baryonic mass
accretion. This while star formation is a global process that, because it
involves many molecular clouds in different evolutionary stages, is likely to
sustain a steady mean rate over times longer than $t_Q$ (Silk \& Norman 2009).
Of course, BH driven outflows may trigger star formation at $>10^2$ M$_\odot$
yr$^{-1}$ (Silk \& Rees 1998). Under these circumstances, AG alone may
not allow the BH mass to keep up, but such star formation episodes are also
much shorter than $t_Q$. Overall, AG at least supports co-eval evolution of
a central supermassive BH and the stellar population of its host galaxy, with
some scatter.
Furthermore, AG allows supermassive BHs ($>10^{10}$ M$_\odot$) to exist in
otherwise normal galaxies, violating the BH mass-velocity dispersion relation
(van den Bosch et al.\ 2012). Also, globular clusters could host AG boosted
intermediate mass BHs of $>10^2$ M$_\odot$. These remain largely undetected
in the radio if outflows weaken their accretion (Sun et al.\ 2013).
Detailed numerical simulations are required to properly assess the interplay
between star and seed BH formation, galaxy merging, gas accretion, stellar
evolution and AG.

BH remnants that are formed only a billion years ago, are not affected
much by AG and must rely on more traditional growth modes.
Conversely, stellar origin BHs formed during the peak in the star formation
history of the universe, at $z=1-2$, are boosted in mass by a factor
$B\approx 10^{4-5}$.
Hence, such quantum grown stelllar BHs can evolve into a much more massive
population and constitute a possible form of dark matter.
In fact, quantum grown BHs can carry about 5 times the baryonic matter content
of a typical galaxy today, as follows.
If it is assumed that there is one stellar origin BH of a few M$_\odot$ in
a thousand stars since $z\sim 1.5$, and a fiducial value of
$B_d\approx 2\times 10^4$ is adopted, then BHs become about 50 times more
significant than the current
stellar mass fraction. This stellar mass is about 10\% of the total baryonic
matter fraction. The resulting dark-to-baryonic ratio of $\sim 5$ is in
agreement with recent Planck data (Planck Collaboration XVI, 2013).

Gravitational interactions (BH-BH and BH-star) may lead to a
complicated spatial distribution of dark matter carried by stellar origin BHs.
One has star-poor/gas-rich dwarf galaxies with a modest/recent
population of stellar BHs, as well as gas-rich spirals and mergers that may
not need AG to feed their central BHs. All the way to passively
evolving massive ellipticals that should be dominated by AG.
The above calls for detailed dynamical calculations beyond the scope
of this paper. In any case, AG of stellar origin BHs yields objects
significantly more massive than a few M$_\odot$. This agrees with
findings of microlensing surveys, which exclude the range of $10^{-7}-10$
M$_\odot$ for dark matter candidates (Alcock 2009).

Obviously, quantum grown stellar BHs are a form of dark matter that appears
only for times later than $t_Q$. If primordial BHs more massive than
$\sim 10^{15}$ g survive to the present time, then these pick up a factor of
$B_d\sim 10^6$. This limits the masses with they can be created at very early
times to conform to observational constraints (Ricotti et al.\ 2008).

There is an exponentially strong dependence of $B_d$ on $t_Q$. This can lead
to a fine-tuning problem when reproducing the dark matter budget, even though
$t_Q$ is large by cosmological standards. It is particularly $t_X$ that
dominates the uncertainty of $t_Q$, given that $A\propto L_f$ has only a $1/3$
power dependence on $N_{\rm BH}(0)$, while $t_X\propto M_X^3$.
With $\Lambda (0)$ well determined, the value of $A$ requires accurate
information on just the current number of macroscopic BHs and
$N_{\rm BH}(0)\approx 10^{19}$ seems reasonable.
This while $M_X$ can only be computed by careful consideration of the
formation, accretion and merging history of primordial BHs in the early
universe (Carr 1975, 2005; Khlopov 2010; Spaans 1997, 2013a).
The factor 10 excursion ($M_X\sim 10m_X$) just captures the notion that large
fluctuations in the BH mass distribution occur during the GUT phase transition.
Nevertheless, the value of $t_Q$ cannot be significantly smaller than $10^9$
years, because that yields too massive BHs today. Conversely, much larger
values suppress quantum foam effects altogether, which can be assessed
observationally.

In any case, this work does not claim that quantum grown stellar and
primordial BHs represent all the dark matter that our universe seems
to need.
Rather, AG may serve as an extra route that naturally emerges and becomes
dominant as time passes.
The above also implies that $B_d>10^{12}$ some 14 billion years from now.
This renders central BHs more massive than their host galaxies, particularly
if (part of) the dark matter halo is accreted during AQ. Therefore, quantum
grown BHs would eventually dominate over dark energy, the latter depending on
BH number only. As such, a big crunch would become the fate of the universe.
\medskip
\noindent The author thanks Tommy de Wilgen for many insightful discussions.
\medskip
\medskip
\noindent {\bf Bibliography}

\noindent Abel, T., Bryan, G. L., \& Norman, M. L.\ 2002, Science, 295, 93

\noindent Alcock, C.\ 2009, ASPC, Vol. 403, in The Variable Universe: A Celebration of Bohdan Paczynski, ed. K. Z. Stanek, p. 71

\noindent Aykutalp, A., Wise, J. H., Meijerink, R., \& Spaans, M.\ 2013, 771, 50

\noindent Aykutalp, A., \& Spaans, M.\ 2011, ApJ, 737, 63

\noindent Bromm, V., \& Loeb, A.\ 2003, ApJ, 596, 34

\noindent Carr, B. J.\ 1975, ApJ, 201, 1

\noindent Carr, B. J.\ 2005, arXiv:astro-ph/0511743

\noindent Di Matteo, T., Springel, V., \& Hernquist, L.\ 2005, Nature, 433, 604

\noindent Fan, X., et al.\ 2001, AJ, 122, 2833

\noindent Ghez, A. M., et al.\ 2008, ApJ, 689, 1044

\noindent Gillessen, S., et al.\ 2009, ApJ, 692, 1075

\noindent Haiman, Z.\ 2006, NewAr, 50, 672

\noindent Hawking, S. W.\ 1974, Nature, 248, 30

\noindent Hopkins, P. F., \& Quataert, E.\ 2010, MNRAS, 407, 1529

\noindent Hopkins, A. M., \& Beacom, J. F.\ 2006, ApJ, 651, 142

\noindent Kennea, J. A., et al.\ 2013, ApJ, 770, L24

\noindent Khlopov, M. Yu., 2010, RAA, 10, 495

\noindent Kim, J.-h., Wise, J. H., Alvarez, M. A., \& Abel, T.\ 2011, ApJ, 738, 54

\noindent Klessen, R. S., Glover, S. C. O., \& Clark, P. C.\ 2012, MNRAS, 421, 3217

\noindent Magorrian, J., et al.\ 1998, AJ, 115, 2285

\noindent Park, K., \& Ricotti, M.\ 2012, ApJ, 747, 9

\noindent Planck Collaboration XVI, 2013, A\&A submitted, arXiv:1303.5076

\noindent Ricotti, M., Ostriker, J. P., \& Mack, K. J.\ 2008, ApJ, 680, 829

\noindent Silk, J., \& Norman, C. A.\ 2009, ApJ, 700, 262

\noindent Silk, J., \& Rees, M. J.\ 1998, A\&A, 331, L1

\noindent Spaans, M.\ 1997, Nuc. Phys. B, 492, 526

\noindent Spaans, M.\ 2013a, Int. J. Mod. Phys. D, Vol.\ 22, 1330022

\noindent Spaans, M.\ 2013b, J. Phys.: Conf. Ser., 410, 012149

\noindent Spaans, M., \& Silk, J.\ 2006, ApJ, 652, 902

\noindent Sun, M.-Y., Lin, Y.-L., Gu, W.-M., Liu, T., Lin, D.-B., \& Lu, J.-F.\ 2013, ApJ in press, arXiv:1308.2869

\noindent van den Bosch, R. et al.\ 2012, Nature, 491, 729

\noindent Wada, K., \& Habe, A.\ 1995, MNRAS, 277, 433

\noindent Wada, K., \& Habe, A.\ 1992, MNRAS, 258, 82

\noindent Wada, K., Papadopoulos, P. P., \& Spaans, M.\ 2009, ApJ, 702, 63

\noindent Wheeler, J. A.\ 1957, Ann. of Phys., 2, 604

\noindent Wheeler, J.A.\ 1987, in Quantum Cosmology, eds. L.Z. Fang \& R. Ruffini (World Scientific, 1987) p. 27-92

\noindent Wise, J. H., Turk, M. J., Norman, M. L., \& Abel, T.\ 2012, ApJ, 745, 50

\noindent Yoshida, N., Omukai, K., \& Hernquist, L.\ 2008, Science, 321, 669

\end